\documentclass{amsart}

\usepackage{amsbsy,amssymb,amscd,amsfonts,latexsym,amstext,delarray,
amsmath,graphicx} 

\newtheorem{thm}{Theorem}[section]
\newtheorem{prop}[thm]{Proposition}
\newtheorem{cor}[thm]{Corollary}
\newtheorem{lem}[thm]{Lemma}

\numberwithin{equation}{section}

\def\bT{{\mathbb T}}

\def\C{{\mathbb C}}

\renewcommand{\H}{{\mathbb H}}
\def\N{{\mathbb N}}
\renewcommand{\P}{{\mathbb P}}
\def\Q{{\mathbb Q}}
\def\Z{{\mathbb Z}}
\def\R{{\mathbb R}}

\def\cA{{\mathcal A}}

\def\cC{{\mathcal C}}

\def\cH{{\mathcal H}}

\def\cR{{\mathcal R}}
\def\cS{{\mathcal S}}

\def\cX{{\mathcal X}}

\def\Tr{{\rm Tr}}
\title{The Ricci flow on noncommutative two-tori}
\author{Tanvir Ahamed Bhuyain}
\author{Matilde Marcolli}
\address{Department of Mathematics  \\
California Institute of Technology \\ 
Pasadena, CA 91125, USA}
\email{tanvirab@caltech.edu}
\email{matilde@caltech.edu}

\begin{document}
\maketitle

\begin{abstract}
In this paper we construct a version of Ricci flow for noncommutative 2-tori,
based on a spectral formulation in terms of the eigenvalues and eigenfunction
of the Laplacian and recent results on the Gauss--Bonnet theorem for 
noncommutative tori.
\end{abstract}

%\tableofcontents

\section{Introduction}

The purpose of this paper is to identify and describe an appropriate 
analog of Hamilton's Ricci flow for the noncommutative two tori, which are
the prototype example of noncommutative manifolds. Our main result is
that we can express the Ricci flow as an equation
\begin{equation}\label{RicciZeta}
\frac{d\lambda}{dt} = \lambda \, 12 \pi \, \zeta_{f^2}(0) ,
\end{equation}
for the eigenvalues and the corresponding eigenfunctions of the 
Laplacian $\triangle' \sim k\triangle k$ considered in \cite{ConTre09},
with the zeta function $\zeta_a(s)= \Tr(a\triangle^{-s})$.
Moreover, \eqref{RicciZeta} admits a reformulation in terms of pseudodifferential
calculus and the properties of the modular operator of the associated
non-unimodular spectral geometry on the noncommutative torus, as in \cite{ConTre09}. This
gives then an equivalent formulation of \eqref{RicciZeta} as
\begin{equation}\label{RicciTau}
\frac{d\lambda}{dt} =  \lambda \, \tau(f^2 {\mathcal R}) ,
\end{equation}
where  $\tau$ is the unique normalized trace \eqref{tautrace} on the noncommutative torus  
and $\cR$ is obtained as
$$ \frac{-1}{12 \pi^2} \cR = 2 \int_0^\infty T(r) b_0 \, r \, dr, $$
where, as in \cite{ConTre09}, one has $\sigma(B_u)(\xi)= b_0(\xi)+b_1(\xi)+b_2(\xi)+\cdots$
the symbol of $B_u=(u{\bf 1} -\triangle')^{-1}$, and $T(r)$ has an explicit expression \eqref{Treq}
involving the symbol. This can then be re-expressed in terms of the modular operator  
$\Delta (a) = e^{-h} a e^h$ with $a \in \cA_{\theta}^{\infty}$ of the non-unimodular geometry
$k\triangle k$ with $k=e^{h/2}$,  as shown in Theorem \ref{RicciRtau}.

\smallskip

\subsection{The Ricci flow and noncommutative geometry}

Hamilton's Ricci flow  \cite{Ham82} on Riemannian manifolds $(M,g)$ is  given
by the evolution equation
\[ \frac{\partial g}{\partial t} = -2Ric(g). \]
The Ricci flow became a fundamental tool in the study of the geometry and topology of manifolds, 
starting with the seminal results of Hamilton \cite{Ham82}, \cite{Ham931}, \cite{Ham932}, 
\cite{Ham95}, and culminating more recently with Perelman's proof of the Poincar\'e conjecture 
\cite{Per02}, \cite{Per031}, \cite{Per032}, see also \cite{And}, \cite{MorTian} and \cite{CaoZhu}, 
\cite{KleiLott},   \cite{Maize}. 

The Ricci flow also found many interesting applications in physics: for instance
in the renormalization group equations of 2-dimensional sigma models \cite{And2}, 
in the evolution of the ADM mass in asymptotically flat spaces \cite{DaiMa}, and also,
recently, in studying the contribution of black holes in Euclidean quantum gravity
\cite{HeadWise}. In view of both the mathematical and the physical interest, and
of the emergence of noncommutative spaces as both an extension of Riemannian
geometry \cite{CoS3}, \cite{Con94} and as a proposed setting for gravity models,
we consider it to be an interesting problem to derive extensions of the Ricci flow
in the setting of noncommutative geometry.

\smallskip

A possible extension of the Ricci flow to
noncommutative spaces was proposed in \cite{Vacaru}, within the
spectral action paradigm for noncommutative geometry (see \cite{ChCo},
\cite{CoS3}), according to which the analog of a (spin) Riemannian manifold in
noncommutative geometry is described by a {\em spectral triple} 
$(\cA,\cH,D)$ as in \cite{CoS3}, with $\cA$ a (noncommutative) algebra
of ``smooth functions", $\cH$ a Hilbert space representation (the
noncommutative analog of square integrable spinors) and $D$ a Dirac
operator. Euclidean gravity on such a noncommutative manifold is then described 
by the {\em spectral action} functional of \cite{ChCo}, which can be thought of as a
modified gravity model. The approach developed in \cite{Vacaru} aims at defining 
an analog of Ricci flow on noncommutative spaces by working directly 
in terms of the spectral action functional.

\smallskip

The approach we follow in this paper is different from the one proposed in \cite{Vacaru}
and it relies on generalizing to the noncommutative setting the behavior of Laplace 
eigenvalues under the Ricci flow on ordinary manifolds, seen as providing a suitable 
{\em spectral version} of the Ricci flow.  

\medskip

\subsection{The Ricci flow and the Laplace spectrum}

The behavior of eigenvalues of the Laplacian
under the Ricci flow on Riemannian manifolds is a subtle and important question: 
it is related, for instance, to Perelman's monotonicity and entropy result in \cite{Per02}, 
which shows that the lowest eigenvalue of the operator $-4\triangle +R$ is non-decreasing
along the Ricci flow, and to other results of a similar flavor in \cite{Cao}, \cite{Cao2},
\cite{Cer07}, \cite{JLi}, \cite{LiMa}.

\smallskip

Here, in order to identify a prototype commutative model of what will happen for 
noncommutative tori, we focus on the result of \cite{Cer07}, where Luca Fabrizio Di Cerbo 
provides an equation for the evolution of Laplace eigenvalues, under the very strong 
assumption of the existence and $\cC^1$-differentiability of the eigenvalues
$\lambda_i(t)$ and eigenfunctions $f_i(t)$, along the deformation $g(t)$ 
of the initial metric determined by the Ricci flow. This strong assumption, which is
in some sense an analog of a Berger and Bando--Urakawa deformation result 
\cite{Ber}, \cite{BaUr}, may be too strong for arbitrary manifolds (see the comments
in \cite{Cer07}).
However, since we are interested in the particular case of the 2-dimensional 
noncommutative tori, we concentrate here only on the commutative case 
of 2-dimensional surfaces, and in particular of the commutative 2-torus.  

\smallskip

The properties of the Ricci flow in 2-dimension were studied in \cite{Ham88} and
\cite{Cho91}, \cite{Ham92}, \cite{HamYau}. As shown in \cite{Ham88}, in 2-dimension
the Ricci flow is related to the gradient flow of the Yamabe problem. For any initial
data, solutions exist for all time. For surfaces of genus $g\geq 1$, in was shown
in \cite{Ham88} that any metric flows to a constant curvature metric and in the genus
zero case any metric with positive Gauss curvature also flows to a constant curvature metric.
The latter assumption in the genus zero case was removed in \cite{Cho91} by showing
that, for any metric on $S^2$, the Gaussian curvature becomes positive in finite time
under the Ricci flow. In particular, in the case of a commutative 2-dimensional torus,
an arbitrary initial metric flows to the flat metric under the Ricci flow, \cite{Ham88}. 

\smallskip

In the 2-dimensional case, under the assumptions described above, 
the equation for the Laplace eigenvalues under
the Ricci flow on a closed surface $M$ is given in Corollary 2.3 of \cite{Cer07} as
\begin{equation}\label{RicciM2}
\frac{d\lambda}{dt} = \lambda\, \int_M f^2 R \, d\mu,
\end{equation}
where $f=f_t$ is the normalized evolving eigenfunction associated to the evolving
eigenvalue $\lambda=\lambda(t)$, with
$$ \int_M f \, d\mu =0 \ \ \ \ \text{ and } \ \ \ \ \int_M f^2 \, d\mu =1. $$

\smallskip

\subsection{The Ricci flow and noncommutative tori}

Our key step in extending \eqref{RicciM2} to the noncommutative case lies in the
formulation of the Gauss--Bonnet theorem for noncommutative tori obtained 
by Connes--Tretkoff in \cite{ConTre09}. 

In the Gauss--Bonnet theorem, the analog of the curvature integral that computes
the Euler characteristic is given in terms of the value at zero of the zeta function of
the Laplacian on the noncommutative torus, and this, in turn, is computed explicitly
in terms of pseudodifferential calculus on the noncommutative space, and proved to
be independent of the Weyl factor as in the classical case. 

In our approach to the Ricci flow, we adapt the same technique to deal with
the values at zero of a more general class of zeta functions associated to the
noncommutative geometry, those of the form $\zeta_a(s)=\Tr(a \, \triangle^{-s})$.

\smallskip

We then conclude the paper by mentioning some possible applications
of this noncommutative extension of the Ricci flow, both in the mathematical
context and in terms of applications to physics.

\section{Ricci flow and zeta functions}

In this section we derive the formula for the Ricci flow on a noncommutative
torus in terms of zeta functions.  We begin by recalling briefly the 
setting of \cite{ConTre09} on noncommutative
tori, variations of the metric within a conformal class, and the Gauss--Bonnet theorem. 

\subsection{Noncommutative tori}

We recall some basic properties of noncommutative tori, which we will need later, 
following \cite{ConnesCR} and \cite{ConTre09}.

Let $\theta$ be an irrational number. Then the irrational rotation $C^*$-algebra $\cA_\theta$ is the universal unital $C^*$-algebra generated by two
unitaries $U,V$ which satisfy the relation
\[ UV = e^{2\pi i\theta}VU, \ \ \ \  \text{ with } \ \ \ U^*U=UU^*=1=V^*V=VV^* . \]
The algebra $\cA_\theta$ is thought of as continuous functions on the 
noncommutative two-torus $\mathbb{T}_\theta^2$, \cite{ConnesCR}. 

A continuous action of $\mathbb{T}^2$, with $(\mathbb{T} =
\mathbb{R}/2\pi \mathbb{Z})$, on $\cA_\theta$ is given by the two-parameter group of automorphism $\{\alpha_s\}$, for $(s\in \mathbb{R}^2)$, determined by
\[ \alpha_s(U^nV^m) = e^{is.(n,m)}U^nV^m . \]
The subalgebra of elements $a\in \cA_\theta$ for which the map $s \mapsto \alpha_s(a)$ is smooth is denoted by $A_\theta^\infty$. Alternatively, as in \cite{ConnesCR} one has
\[ \cA_\theta^\infty = \{\sum_{m,n\in \mathbb{Z}} a_{m,n}U^mV^n:(|m|^k|n|^q|a_{m,n}|) \mbox{ is bounded for any positive } k,q\} . \]

The corresponding derivations are given by
\[ \delta_1(U) = U,\mbox{  } \delta_1(V) = 0 , \]
\[ \delta_2(U) = 0,\mbox{  } \delta_2(V) = V . \]
These derivations are viewed as the analogs to the differential operators 
$-i \partial/\partial x$, $-i \partial/\partial y$ of the
classical case. We also have 
\[ \delta_j(a^*) = -\delta_j(a)^*, \mbox{  } j=1,2 . \]

For irrational $\theta$, there is a unique normalized trace 
$\tau$ on $\cA_\theta$ given by
\begin{equation}\label{tautrace}
 \tau(U^nV^m) = 0 \ \ \ \text{ if } \ \  (n,m) \neq (0,0), \ \ \ \ \text{ and } \ \ \ \tau(1) = 1 .
\end{equation} 
This satisfies the analog of the classical integration by parts formula
\[ \tau \circ \delta_j = 0, \mbox{     } \tau(a\delta_j(b)) = - \tau(\delta_j(a)b), \mbox{   } \forall a,b\in \cA_\theta^\infty, \mbox{  } j=1,2 . \]
More details on the structure of the noncommutative torus can be found, for example, in
\cite{Con94}, \cite{Var06}.

\medskip

It is customary to consider a Hilbert space $H_0$ constructed
as the completion of $\cA_\theta$ with respect to the inner product
\[ \langle a,b\rangle = \tau(b^*a), \ \ \ \ a,b\in \cA_\theta . \]
With respect to this inner product, $\delta_1, \delta_2$ are formally self adjoint unbounded operators on $H_0$. One can then add a complex structure by
defining
\[ \partial = \delta_1 + i \delta_2, \ \ \ \  \partial^* = \delta_1 - i \delta_2 . \]
The operator $\partial$ is then an unbounded operator on $H_0$ and $\partial^*$ 
is its formal adjoint with respect to the inner product determined by $\tau$.

The information on the conformal structure is contained in the positive Hochschild two-cocycle described in \cite{ConCun88} and \cite{Con94},
\[ \psi(a,b,c) = -\tau(a\partial b\partial^*c), \ \ \ \ \  a,b,c\in \cA_\theta^\infty . \]
The analog of the classical space of $(1,0)$-forms is the unitary bimodule 
$H^{(1,0)}$ on $\cA_\theta$ obtained by the Hilbert space completion of the
space of finite sums $\sum a \partial b$, where $a,b \in \cA_\theta^\infty$, with respect to the inner product described above. We consider
$\partial$ to be an unbounded operator from $H_0$ to $H^{(1,0)}$. 
The Laplacian $\triangle$ on the functions on $\mathbb{T}_\theta^2$  is then defined
as
\begin{equation}\label{Laplacian}
 \triangle = \partial^* \partial = \delta_1^2 + \delta_2^2 . 
\end{equation} 

\subsection{Variation of the metric and the Laplacian}

We continue with our brief review of the notions of \cite{ConTre09} that we will need to use in
the following.

We vary inside the conformal class of a metric by choosing a self adjoint 
element $h \in A_\theta^\infty$  and defining the positive
linear functional
\begin{equation}\label{varphih}
 \varphi(a) = \tau(ae^{-h}), \ \ \ \ \  a\in \cA_\theta . 
\end{equation}

One constructs then a unitary left module $H_\varphi$ on $\cA_\theta$ 
by completing $\cA_\theta$ with respect to the inner product
\[ \langle a,b\rangle_\varphi = \varphi(b^*a), \ \ \ \   a,b\in A_\theta . \]

One then considers $\partial_\varphi$, which is the same operator as $\partial$ but viewed as an unbounded operator from $H_\varphi$ to $H^{(1,0)}$, and one
constructs the modified Laplacian $\triangle'$ as a positive unbounded 
operator acting in $H_\varphi$ given by 
\begin{equation}\label{Laplacian2}
 \triangle' = \partial_\varphi^*\partial_\varphi . 
\end{equation} 

One then has the following result.

\begin{lem}\label{kDelta} {\rm (Connes--Tretkoff \cite{ConTre09})}
The operator $\triangle'$ acting on $H_\varphi$ is anti-unitarily equivalent to the positive 
unbounded operator $k\triangle k$
acting on $H_0$, where $k=e^{h/2}\in \cA_\theta$ acts by left multiplication on $H_0$.
\end{lem}

We refer the reader to \cite{ConTre09} for more details. 

\smallskip

By Lemma \ref{kDelta}, one sees that $\triangle'$ and $k\triangle k$  have the same eigenvalues. 
Thus, we will be considering an analog of the Ricci flow equation \eqref{RicciM2}
obtained by Di Cerbo in \cite{Cer07}, in the form of a time evolution equation for the operator
$k\triangle k$ involving its eigenvalues. 
This will give us an evolution of $k=e^{h/2}$ and subsequently an
evolution for $h \in A_\theta^\infty$, which parametrizes the metric 
$\varphi$  (through the inner product induced by $\varphi$),  with 
$\varphi(a) = \tau (ae^{-h})$ as above.

\subsection{Zeta functions for noncommutative tori}

For a classical (commutative) closed 2-dimensional surface (in particular the two torus), 
the spectral zeta function of the 
Laplacian $\triangle$ is given by
\begin{equation}\label{zetaT2}
 \zeta(s) = \sum_j \lambda_j^{-s} = \Tr(\triangle^{-s}), \ \ \ \ \ \text{ for } \ {\rm Re}(s) > 1, 
\end{equation} 
where $\lambda_j$ are the eigenvalues of $\triangle$ and we have assumed that 
${\rm Ker}(\triangle)=0$. 

The function $\zeta(s)$ has a meromorphic continuation to the origin $s=0$, 
where it has no pole, and we have (still assuming ${\rm Ker}(\triangle)=0$)
\begin{equation}\label{zeta0int}
 \zeta(0) = \frac{1}{12\pi}\int_M R \, d\mu =\frac{1}{6} \chi(M) , 
\end{equation}
with $\chi(M)$ the topological Euler characteristic (by Gauss--Bonnet theorem), 
which vanishes in the case of the 2-torus $\mathbb{T}^2$. 

%%%% add some comment here: Gauss--Bonnet
 
According to a general philosophy developed recently in \cite{Corn},
\cite{CornMa1},  \cite{CornMa2},  a good ``set of coordinates" on a 
noncommutative geometry is provided by a {\em family of zeta functions},
which generalize the zeta function of the Laplacian $\triangle$ (or of a 
Dirac operator $D$) by including elements of the algebra $\cA$, in the
form
\begin{equation}\label{genZeta}
\zeta_a(s) = \Tr(a \, \triangle^{-s}) .
\end{equation}
In the case of an ordinary Riemannian manifold, it is proved in \cite{Corn}
that a similar family of zeta functions reconstructs the manifold up to isometry.
Here we will use specific zeta functions in this family to extend the integration
\eqref{zeta0int} to an expression that recovers, for ordinary (commutative)
2-tori, the eigenvalue equation for the Ricci flow of \cite{Cer07} and continues
to make sense for all the noncommutative 2-tori. 

\smallskip

\begin{prop}\label{classT2zeta}
In the case of the closed 2-dimensional surface $M$, let $f$ be an element in the
smooth subalgebra $C^\infty(M)$ and consider the zeta function 
$\zeta_f(s)=\Tr( f \triangle^{-s})$ as above. Then $\zeta_f(s)$ has a
meromorphic continuation at $s=0$ where it satisfies
\begin{equation}\label{zetaf0}
\zeta_f(0)= \frac{1}{12\pi} \int_M f \, R \, d\mu .
\end{equation}
\end{prop}

\proof
We can work under the assumptions of Lemma 1.2.1 of \cite{Gilkey}, see also (1.12.14)
of \cite{Gilkey}, since $\Tr( f e^{-t\triangle})$ is admissible in the sense specified
there. Then $\zeta_f(s)=\Tr( f \triangle^{-s})$ can be written via Mellin transform as
$$ \Tr( f \triangle^{-s}) = \frac{1}{\Gamma(s)} \int_0^\infty \Tr( f e^{-t\triangle}) \, t^{s-1}dt. $$
The function $\Gamma(s) \zeta_f(s)$ has a meromorphic extension to $\C$, with residues
given in terms of the coefficients $a_n$ of the heat-kernel expansion, see  Theorem 1.12.2 
of \cite{Gilkey}. In particular, $\zeta_f(s)$ is regular at $s=0$. As in \cite{Gilkey}, we have
in general for $m=\dim M$, 
$$ \Tr( f e^{-t \triangle}) \sim \sum_n t^{(n-m)/2} \int_M f \, a_n(\triangle)\, d\mu, $$
so that, combining this with what we already know from
\eqref{zeta0int}, we obtain as the value $\zeta_f(0)$ 
the integration \eqref{zetaf0}.
\endproof

\smallskip

We can then reformulate the equation \eqref{RicciM2} in terms of the zeta functions
as follows.

\begin{cor}\label{RicciT2zeta}
Let $M$ be a closed 2-dimensional surface. The Ricci flow equation \eqref{RicciM2}
for the eigenvalues of the Laplacian can be rewritten as
\begin{equation}\label{RicciM2zeta}
\frac{d\lambda}{dt} = \lambda\, 12\pi \, \zeta_{f^2}(0),
\end{equation}
where $\lambda$ is an evolving eigenvalue and $f$ the corresponding normalized
evolving eigenfunction.
\end{cor}

\proof This follows immediately from the previous Proposition.
\endproof

Now one can simply observe that the right-hand-side of \eqref{RicciM2zeta}
continues to make sense when passing from commutative to noncommutative tori,
hence it can be taken as the appropriate extension of the right-hand-side 
of \eqref{RicciM2} in this context. 

\smallskip

Thus, we conclude that a good analog of Ricci flow on noncommutative
tori is provided by the equation \eqref{RicciZeta}.

\smallskip

In the following, we obtain a more explicit expression for the
right-hand-side of \eqref{RicciM2zeta} in the case of noncommutative
tori, in terms of pseudodifferential calculus and the modular operator,
as in \cite{ConTre09}.

\medskip

\section{Value at the origin}\label{valuesec}

We first express the value $\zeta_a(0)$ in terms of pseudodifferential
calculus, as in \cite{ConTre09}.

\begin{prop}\label{zetaf0b2}
Consider a pseudodifferential operator with symbol 
\begin{equation}\label{symbolb}
 \sum_{j \geq 0} \, b_j(\xi_1,\xi_2), 
\end{equation}
which is an approximation to the inverse of the operator $(\triangle'+1)$ in the space
of symbols, with $(\xi_1,\xi_2)$ the canonical coordinate system of $\mathbb{R}^2$. Here $b_j(\xi_1, \xi_2)$ is homogeneous of order $-2-j$.
Then 
\begin{equation}\label{zeta0symb}
\zeta_a(0)= - \int \tau(a \, b_2(\xi_1,\xi_2)) \, d \xi_1 \, d\xi_2 = - \tau \Big(a \int  \, b_2(\xi_1,\xi_2) \, d \xi_1 \, d\xi_2 \Big) .
\end{equation}
\end{prop}

\proof  The pseudodifferential calculus for noncommutative tori associates to 
a symbol $\sigma(\xi)=\sum_{|j|\leq n} a_j \xi_1^{j_1} \xi_2^{j_2}$,
with $a_j\in \cA^\infty_\theta$ a differential operator 
$P_\sigma=\sum_{|j|\leq n} a_j \delta_1^{j_1} \delta_2^{j_2}$ and, more generally,
a pseudodifferential operator to an arbitrary symbol $\sigma\in \cS=\cup_{n\in \Z}\cS_n$, 
where $\cS_n$ denotes the symbols of order $n$ as in Definition 4.1
of \cite{ConTre09}.

For the Laplacian $\triangle' \sim k\triangle k$, under the assumption that
${\rm Ker}(\triangle')=0$, we can use Mellin transform to get 
$$  \zeta(s)= \Tr(\triangle'^{-s}) = \frac{1}{\Gamma(s)} \int_0^\infty \Tr(e^{-t\triangle'}) \, t^{s-1}\, dt $$
Using symbols, one can express the zeta function  as an integration 
$$ \zeta(s)= \frac{1}{\Gamma(s)} \int_0^\infty \int \tau(\sigma(e^{-t\triangle'})(\xi)) \,\, 
t^{s-1} \, d\xi \, dt.  $$
If ${\rm Ker}(\triangle)\neq 0$, one replaces $\Tr(e^{-t\triangle'})$ with $\Tr^+(e^{-t\triangle'})=
\Tr(e^{-t\triangle'}) - \dim {\rm Ker}(\triangle')$.

We can proceed in a similar way for the zeta functions $\zeta_a(s)= \Tr(a\triangle'^{-s})$.
The meromorphic extension of the function $\zeta_a(s)$ to $\C$ is equivalent 
(\cite{Scott} \S 3.3.2.1-3.3.2.2) to the existence of an asymptotic expansion for the 
resolvent traces $\Tr(a \partial^p_\lambda (\triangle'-\lambda)^{-1})$. Thus, by
$$ e^{-t \triangle'} = \int_C e^{-\lambda} (t\triangle'-\lambda)^{-1} d\lambda =
t^{-p-1} \int_C e^{-\lambda} \partial^p_{\lambda t^{-1}} (\triangle'-\lambda t^{-1})^{-1} d\lambda, $$
the generalized heat trace $\Tr(a e^{-t\triangle'})$ also has an asymptotic 
expansion in the limit $t \rightarrow 0^+$, which we can write as
\begin{equation}\label{Traheat}
 \Tr(a e^{-t\triangle'})\sim t^{-1} \sum_{n=0}^\infty B_{2n}(a,\triangle') t^n . 
\end{equation} 

The zeta function $\zeta_a(s)$ and the generalized heat trace are related by the
general heat-zeta transition formulae (\cite{Scott}, \S 3.3.3.2) using Mellin transform
\begin{equation}\label{Mellin}
\zeta_a(s)= \Tr(a\triangle'^{-s})= \frac{1}{\Gamma(s)} \int_0^\infty \Tr(a\, e^{-t\triangle'})t^{s-1}dt,
\end{equation}
and 
\begin{equation}\label{MellinInv}
\Tr(a e^{-t\triangle'}) =\frac{1}{2\pi i} \int_{{\rm Re}(s)=c} t^{-s} \, \zeta_a(s) \Gamma(s) \, ds .
\end{equation}
Here one replaces $\Tr(a e^{-t\triangle'})$ with $\Tr^+(a e^{-t\triangle'})$ if triviality of the kernel of $\triangle'$ is not assumed.

The very general Grubb--Seeley transition formulae (\cite{Scott} \S 3.3.3.2) relate an
asymptotic expansion 
$$ \epsilon(t) \sim \sum_{j=0}^\infty \sum_{\ell=0}^{n_j} a_{j,\ell} t^{\beta_j} \log^\ell(t) $$
for an increasing sequence of $0< \beta_j\to \infty$ and with $n_j\in \N$, to the meromorphic
structure of the Mellin transform $f(z) = \int_0^\infty t^{z-1} \epsilon(t)\, dt$, by
$$ 
f(z) \sim \sum_{j=0}^\infty \sum_{\ell=0}^{n_j} \frac{(-1)^\ell\, \ell! \, a_{j,\ell}}{(z+\beta_j)^{\ell+1}}. 
$$
When applied to the particular case (and simpler form) of the asymptotic expansion 
\eqref{Traheat}, considering the fact that the Gamma function has a simple pole at zero, we get
$$ \zeta_a(0) = B_2(a,\triangle'). $$
Equivalently, using the symbols calculus, we have
$$  \frac{1}{\Gamma(s)} \int_0^\infty \Tr(a\, e^{-t\triangle'})t^{s-1}dt
 = \frac{1}{\Gamma(s)} \int_0^\infty \int \tau( a \sigma(e^{-t\triangle'})(\xi)) t^{s-1} \, d\xi \, dt , $$
so that
$$  \zeta_a(0) = {\rm Res}_{s=0} \int_0^\infty \int \tau(a\, \sigma(e^{-t\triangle'})(\xi)) \, 
t^{s-1}\, d\xi\, dt, $$
where, for $t\to 0^+$,
$$ \int \tau(a\, \sigma(e^{-t\triangle'})(\xi))\, d\xi \sim t^{-1} 
\sum_{n=0}^\infty B_{2n}(a,\triangle') t^n . $$
Using again the Cauchy integral formula 
$$ e^{-t \triangle'}= \frac{1}{2\pi i} \int_C e^{-t\lambda} (\triangle' - \lambda)^{-1} \, d\lambda $$
and approximating the inverse $(\triangle' - \lambda)^{-1}$ by a pseudodifferential operator
$B_\lambda$ with symbol \eqref{symbolb}, as in \cite{ConTre09} and \cite{ZadKha10}, one
then finds
$$ 
B_2(a,\triangle')= \frac{1}{2\pi i} \int_C e^{-\lambda} \, \tau( a b_2(\xi,\lambda))\, d\lambda\, d \xi ,
$$
and, by the same homogeneity argument used in \cite{ConTre09} and \cite{ZadKha10}, one then
finds, as in \eqref{zeta0symb} that $\zeta_a(0) = - \tau ( a \, \int b_2(\xi) \, d\xi )$. 
\endproof
 
We can then rewrite the above result, using the following computation as in \cite{ConTre09}.

\begin{lem}\label{intb2Tr}
Consider the pseudodifferential operator with symbol \eqref{symbolb} as above. Then
\begin{equation}\label{intb2Treq}
\int  \, b_2(\xi_1,\xi_2) \, d \xi_1 \, d\xi_2 =  2 \pi \int_0^\infty  \, T(r) \, b_0 \, r \, dr,
\end{equation}
where $b_0 = (k^2 r^2 + 1)^{-1}$ and $T(r)$ is given by the explicit expression (using Einstein's summation notation)
\begin{align}\label{Treq} \nonumber
T(r) & = - b_0k\delta_i \delta_i(k) + 3r^2k^3b_0^2\delta_i\delta_i(k) + 4r^2k^2b_0^2\delta_i(k)\delta_i(k) + r^2k^2b_0^2\delta_i\delta_i(k)k \\ \nonumber
& - 2r^4k^5b_0^3\delta_i\delta_i(k) - 4r^4k^4b_0^3\delta_i(k)\delta_i(k) - 2r^4k^4b_0^3\delta_i\delta_i(k)k \\
\nonumber
& + r^2kb_0\delta_i(k)b_0(4k\delta_i(k)+2\delta_i(k)k) - 2r^4kb_0\delta_i(k)k^2b_0^2(k\delta_i(k) + \delta_i(k)k) \\ \nonumber
& - r^4k^3b_0^2\delta_i(k)b_0(8k\delta_i(k)+6\delta_i(k)k) + 2r^6k^3b_0^2\delta_i(k)k^2b_0^2(k\delta_i(k)+\delta_i(k)k) \\ 
& - r^4k^2b_0^2\delta_i(k)kb_0(6k\delta_i(k)+4\delta_i(k)k) + 2r^6k^2b_0^2\delta_i(k)k^3b_0^2(k\delta_i(k)+\delta_i(k)k) \\  \nonumber
& + 4r^6k^5b_0^3\delta_i(k)b_0(k\delta_i(k)+\delta_i(k)k) + 4r^6k^4b_0^3\delta_i(k)kb_0(k\delta_i(k)+\delta_i(k)k). \nonumber
\end{align}
\end{lem} 
 
\proof 
In \cite{ConTre09} it is shown that 
\begin{equation}\label{intb2}
 \int  \, b_2(\xi_1,\xi_2) \, d \xi_1 \, d\xi_2 =  \int  \, S(\xi_1,\xi_2)b_0 \, d \xi_1 \, d\xi_2, 
\end{equation} 
where $b_0 = (k^2 \xi_1^2 +k^2  \xi_2^2 + 1)^{-1}$ and $S(\xi_1,\xi_2)$ is given
by the explicit expression \eqref{Sxi1xi2} reported in the Appendix 
below.  After passing to polar coordinates in \eqref{Sxi1xi2} we obtain \eqref{Treq}.
\endproof

\smallskip 

\subsection{The modular operator} For consistency, we will be following 
here the somewhat unfortunate choice
of notation adopted in \cite{ConTre09}, according to which we will
continue to denote the {\em Laplacian} by $\triangle$, while at the same time
using the very similar notation $\Delta$ for the {\em modular operator}. 

The latter is defined by 
\begin{equation}\label{Deltah}
\Delta (x) = e^{-h} x e^h,  \ \ \ \    \text{ for }  \ \  x \in \cA_{\theta}^{\infty},
\end{equation}
for $h$ as in \eqref{varphih}.

We recall the following result on the modular operator $\Delta$.

\begin{lem}\label{lemDm} {\rm (Connes--Tretkoff \cite{ConTre09})}
For every element $\rho$ of $\cA_{\theta}^{\infty}$ and every
non-negative integer $m$ one has
\begin{equation}\label{Dmrho}
\int_0^{\infty} \frac{k^{2m+2} \, u^m }{ (k^2 \, u + 1)^{m+1}}
\,\rho\, \frac{1}{(k^2 \, u+1)} \, du = {\mathcal D}_m (\rho) ,
\end{equation}
where 
\begin{equation} \label{Dmint}
{\mathcal D}_m  = \int_0^{\infty} \frac{x^m }{
(x+1)^{m+1}}\frac{1}{(x \Delta + 1)} \, dx . 
\end{equation}
\end{lem}

We refer the reader to \cite{ConTre09} for more details. In the same way, we
obtain the following variant.

\begin{prop}\label{propDm} 
For $\lambda, \rho \in \cA_\theta^\infty$ and $w,y,v \in \mathbb{Z}_{>0}$, 
\begin{equation}\label{Dwvy}
\int_0^{\infty} \frac{1}{(k^2u+1)^v}\lambda \frac{k^{2w}u^{(w-1)}}{(k^2u+1)^y} \rho \frac{1}{(k^2u+1)}
du = D_{w,v,y}(\lambda,\rho) 
\end{equation}
where 
\begin{equation}\label{Dwvy2}
D_{w,v,y}(\lambda,\rho) = \int_{0}^{\infty} \, \Big[\frac{1}{(x\Delta^{-1}+1)^v}(\lambda)\Big] \,  \frac{x^{w-1}}{(x+1)^y} \,
\Big[\frac{1}{(x\Delta+1)}(\rho)\Big] \, dx .
\end{equation}
\end{prop} 
 
\proof Using $k = e^{h/2}$ and the change of variable $u = e^s$, we get
$$  \int_0^{\infty} \frac{1}{(k^2u+1)^v}\lambda \frac{k^{2w}u^{(w-1)}}{(k^2u+1)^y} \rho \frac{1}{(k^2u+1)} du $$
$$  = \int_{-\infty}^{\infty} \frac{1}{(e^{(s+h)}+1)^v}\lambda \frac{e^{hw+s(w-1)}}{(e^{(s+h)}+1)^y} \rho \frac{e^s}{(e^{(s+h)}+1)} ds $$
$$  = \int_{-\infty}^{\infty} \frac{e^{(s+h)/2}}{(e^{(s+h)}+1)^v}\Delta^{1/2}(\lambda) \frac{e^{(s+h)(w-1)}}{(e^{(s+h)}+1)^y} \Delta^{-1/2}(\rho)
\frac{e^{(s+h)/2}}{(e^{(s+h)}+1)} ds $$
$$ = \int_{-\infty}^{\infty} \frac{1}{(e^{(s+h)}+1)^{(v-1)}} \Upsilon(h,s) \,  
\Delta^{1/2}(\lambda) \frac{e^{(s+h)(w-1)}}{(e^{(s+h)}+1)^y} \Delta^{-1/2}(\rho) 
 \Upsilon(h,s) \,  ds , $$
where
$$ \Upsilon(h,s) = \int_{-\infty}^{\infty}\frac{e^{it(s+h)}}{(e^{\pi t}+e^{-\pi t})} dt . $$
We then further write the above as
$$ = \int_{-\infty}^{\infty} \frac{1}{(e^{(s+h)}+1)^{(v-1)}} \,
\Xi_{1, (h,s)}(\lambda) \,  \frac{e^{(s+h)(w-1)}}{(e^{(s+h)}+1)^y} \,  \Lambda_{ (h,s)}(\rho)  \,ds, $$
where
$$ \Xi_{1, (h,s)}(\lambda)=  \int_{-\infty}^{\infty} \Delta^{1/2-it}(\lambda) \frac{e^{it(s+h)}}{(e^{\pi
t}+e^{-\pi t})} dt  $$
$$ \Lambda_{(h,s)}(\rho)= \int_{-\infty}^{\infty}\frac{e^{it(s+h)}}{(e^{\pi t}+e^{-\pi t})}
\Delta^{-1/2+it}(\rho)\, dt . $$
We can then proceed by induction to obtain
$$ = \int_{-\infty}^{\infty}  \Xi_{v, (h,s)}(\lambda) \, \frac{e^{(s+h)(w-1)}}{(e^{(s+h)}+1)^y}  \, 
\Lambda_{ (h,s)}(\rho) \,  ds, $$
with $\Lambda_{ (h,s)}(\rho)$ as above and with
$$ \Xi_{v, (h,s)}(\lambda) = \prod_{j=1}^v \int_{-\infty}^{\infty} \Delta^{1/2-it_j}(\lambda) \frac{e^{it_j(s+h)}}{(e^{\pi t_j}+e^{-\pi t_j})} dt_j . $$
We further write the above as
\begin{align*}
= & \int_{-\infty}^{\infty}  \int_{-\infty}^{\infty} \cdots \int_{-\infty}^{\infty}\Big( \prod_{j=1}^v \Delta^{1/2-it_j}(\lambda)
\frac{e^{it_j h}}{(e^{\pi t_j}+e^{-\pi t_j})}\Big) \\
& \ \  \times\Big( \int_{-\infty}^{\infty}
\frac{e^{(s+h)(w-(v+1)/2)}((e^{its}\prod_{j=1}^ve^{it_js}))}{(e^{(s+h)}+1)^y} ds \Big) \\
& \ \  \times \frac{e^{it h}}{(e^{\pi t}+e^{-\pi t})} \Delta^{-1/2+it}(\rho)
dt \, dt_1 \cdots dt_v .
\end{align*}
We denote by $F_{w,v,y}$ the Fourier transform of the function
$$ h_{w,v,y}(s) = \frac{e^{s(w-(v+1)/2)}}{(e^{s}+1)^y}. $$
The expression above then becomes
\begin{align*}
= &  \int_{-\infty}^{\infty}  \int_{-\infty}^{\infty} \cdots \int_{-\infty}^{\infty}\Big( \prod_{j=1}^v \Delta^{1/2-it_j}(\lambda) \frac{1}{(e^{\pi
t_j}+e^{-\pi t_j})}\Big) \\
& \ \  \times F_{w,v,y}  
 \Big(t+ \sum_{j=1}^v t_j\Big)  \frac{1}{(e^{\pi t}+e^{-\pi t})} \Delta^{-1/2+it}(\rho) dt \, dt_1 \cdots dt_v  
\end{align*}
$$ = \int_{0}^{\infty} \Big[\Big(\frac{x^{1/2}}{x\Delta^{-1}+1}\Big)^v(\lambda)\Big] \frac{x^{(w-(v+1)/2)}}{(x+1)^y}
\Big[ \Big(\frac{x^{1/2}}{x\Delta+1})(\rho \Big)\Big] \frac{dx}{x} $$
$$ = \int_{0}^{\infty} \Big[\frac{1}{(x\Delta^{-1}+1)^v}(\lambda)\Big] \frac{x^{w-1}}{(x+1)^y} \Big[\frac{1}{(x\Delta+1)}(\rho)\Big] dx . $$
This is the expression $D_{w,v,y}(\lambda,\rho)$ of \eqref{Dwvy2}.
\endproof 

We can now conclude the argument and give an expression for the Ricci
flow in terms of these properties of the modular operator.

\begin{thm}\label{RicciRtau}
The Ricci flow on the noncommutative torus is given by
\begin{equation}\label{RicciRtaueq}
\frac{d\lambda}{dt} = \lambda \, 12 \pi \, \zeta_{g^2}(0) = \lambda \, \tau(f^2 {\mathcal R}) ,
\end{equation}
where $\lambda$ and $f$ represent (respectively) the evolving eigenvalues and the 
corresponding evolving normalized eigenfunctions of the Laplacian 
$\triangle' \sim k\triangle k$ and
\begin{align*}\label{cReq} 
\frac{-1}{12\pi^2}{\mathcal R} & = \sum_{i=1,2} ( \,
- k^{-1}{\mathcal D}_0(\delta_i^2(k)) 
+ 3k^{-1}{\mathcal D}_1 (\delta_i^2(k)) \\ 
& + 4k^{-2}{\mathcal D}_1 (\delta_i(k)^2) + k^{-2}{\mathcal D}_1 (\delta_i^2(k)k) 
- 2k^{-1}{\mathcal D}_2 (\delta_i^2(k)) \\ 
& - 4k^{-2}{\mathcal D}_2 (\delta_i(k)^2) - 2k^{-2}{\mathcal D}_2 (\delta_i^2(k)k) \\
& + kD_{2,1,1}(\delta_i(k)k^{-4},4k\delta_i(k)  + 2\delta_i(k)k) \\
& - 2kD_{3,1,2}(\delta_i(k)k^{-4},k\delta_i(k) 
+ \delta_i(k)k)  \\ 
& - k^3D_{3,2,1}(\delta_i(k)k^{-6},8k\delta_i(k)
+ 6\delta_i(k)k) \\
& + 2k^3D_{4,2,2}(\delta_i(k)k^{-6},k\delta_i(k)
+\delta_i(k)k) \\ 
& - k^2 D_{3,2,1}(\delta_i(k)k^{-5},6k\delta_i(k)
+ 4\delta_i(k)k) \\ 
& + 2k^2 D_{4,2,2}(\delta_i(k)k^{-5},k\delta_i(k)
+ \delta_i(k)k) \\ 
& + 4k^5D_{4,3,1}(\delta_i(k)k^{-8},k\delta_i(k)
+ \delta_i(k)k) \\ 
& + 4k^4D_{4,3,1}(\delta_i(k)k^{-7},k\delta_i(k)+\delta_i(k)k) \, ),
\end{align*} 
with  $k = e^{h/2}$, $h = h^{*} \in \cA_\theta^\infty$,  and  $\Delta$ 
defined by $ \Delta(x) = e^{-h}xe^h$ 
for $x \in \cA_\theta^\infty$, and with 
\[ {\mathcal D}_m = \int_0^{\infty} \frac{x^m }{
(x+1)^{m+1}}\frac{1}{(x\Delta + 1)} \, dx \, , \]
\[ D_{w,v,y}(\gamma,\rho) = \int_{0}^{\infty} \, 
\Big[\frac{1}{(x\Delta^{-1}+1)^v}(\gamma)\Big] \,  \frac{x^{w-1}}{(x+1)^y} \,
\Big[\frac{1}{(x\Delta+1)}(\rho)\Big] \, dx,  \]
for $\gamma, \rho \in \cA_\theta^\infty$.
\end{thm}

\proof In Lemma \ref{intb2Tr} we have 
$$ \zeta_a(0) = \frac{1}{12\pi} \, \tau(a {\mathcal R}), $$
with
$$
 \frac{-1}{12\pi^2}{\mathcal R} =  2 \int_0^\infty  \, T(r) \, b_0 \, r \, dr = 
\int_0^\infty  \, T(r) \, b_0 \, 2 r \, dr. $$
Using Lemma \ref{lemDm} and Proposition \ref{propDm}, with the change 
of variable $u=r^2$, we can rewrite the explicit expression \eqref{Treq}  for $T(r)$ 
and obtain the result as stated.   
\endproof

\medskip

\section{Noncommutative black holes}

We give here an illustration of a possible physical application 
of the Ricci flow on noncommutative tori discussed in the
previous sections. 

\smallskip

\subsection{The BTZ black hole}

The Ba\~{n}ados--Teitelboim--Zanelli (BTZ) black hole \cite{BTZ} is a black
hole in $(2+1)$-dimensional gravity. It is an asymptotically AdS$_{2+1}$
space obtained as a global quotient of anti-de Sitter space by a discrete
group of isometries $\Gamma \subset SO(2,2)$ generated by a single
loxodromic element. In Euclidean gravity, the Euclidean version of the BTZ
black hole is given by a quotient $\cX_q=\H^3/(q^\Z)$ of 3-dimensional
real hyperbolic space $\H^3$ (the Euclidean version of AdS$_{2+1}$)
by a subgroup $\Gamma=q^\Z$ of ${\rm PSL}(2,\C)$ generated by an element $q\in \C^*$
with $|q|<1$, acting on $\H^3$ by $(z,y) \mapsto (qz, |q|y)$ in the realization
of $\H^3$ as the upper half space $\C \times \R^*_+$, see \cite{BKSW},
\cite{Kra}, \cite{ManMar}.  The space $\cX_q$ obtained in this way
describes a spinning black hole whenever $q$ is not purely real. 
Topologically $\cX_q$ is a solid torus, endowed with
a hyperbolic metric with conformal boundary at infinity given by the 
Tate-uniformized elliptic curve $E_q(\C)=\C^*/(q^\Z)$. In other words,
the compactification $\overline{\cX}_q=\cX_q\cup E_q(\C)$ is the
quotient $\Omega_\Gamma /\Gamma$ of the action of
$\Gamma=q^\Z$ on the domain of discontinuity $\Omega_\Gamma \subset \H^3\cup \P^1(\C)$,
that is, the subset of the compactified $\overline{\H^3}=\H^3\cup \P^1(\C)$ 
on which $\Gamma$ acts freely and properly discontinuously. This is the
complement of the limit set $\Lambda_\Gamma=\{0,\infty\} \subset \P^1(\C)$. 

\smallskip

In recent years, noncommutative deformations of the BTZ black hole
were considered in the context of string theory compactifications
and of various proposals for noncommutative models of gravity, see
for instance \cite{BDRS}, \cite{DGS}, \cite{EeLee}.  

\smallskip
\subsection{From elliptic curves to noncommutative tori}

We consider here a very natural noncommutative deformation 
of the Euclidean BTZ black hole, based on considering a degenerating
family of BTZ black holes, where the uniformization parameter 
$q$ tends to a point on the unit circle with irrational angle 
$q\to e^{2\pi i \theta}$, with $\theta\in \R \smallsetminus \Q$. 
This has the effect of replacing the elliptic curve $E_q(\C)=\C^*/(q^\Z)$
with a ``bad quotient" described by (the suspension of) a 
noncommutative torus with algebra of functions
$C(S^1)\rtimes_\theta \Z=\cA_\theta$. The suspension corresponds to
the fact that the group $\Gamma = e^{2\pi i \theta \Z}$ now acts trivially
on the radial direction in $\C^*=S^1\times R^*_+$ so that the crossed
product algebra describing the quotient is given by
$\cA_\theta \otimes C_0(\R^*_+)$. 

Notice how in this case, while one
thinks of the noncommutative torus itself as a 2-dimensional geometry,
in the degeneration of the elliptic curve $E_q(\C)$ when $q\to e^{2\pi i \theta}$,
it accounts only for a 1-dimensional geometry, which is the ``bad quotient" of a
circle by the action of $\Gamma=e^{2\pi i\theta \Z} \simeq \Z$ by irrational
rotations. One is left with a commutative radial direction $\R^*_+$ which
is not affected by the quotient, since it is acted upon trivially.  From the
point of view of Ricci flow, this now allows for more interesting behaviors
than in the original commutative case, intuitively because a one-dimensional
geometry has been replaced in the limit by a richer two-dimensional (but
noncommutative) geometry. We discuss this briefly in \S \ref{ncblacksec} below.

As the horizon $E_q(\C)$ degenerates
in this way to a noncommutative torus $\bT^2_\theta$, described in terms of
the algebra $\cA_\theta \otimes C_0(\R^*_+)$ above, the bulk space $\cX_q$ also
is correspondingly deformed to a noncommutative space. In the limit
$q\to e^{2\pi i \theta}$ the action on $\H^3$ becomes $(z,y) \mapsto (e^{2\pi i \theta}z,y)$.
The resulting ``bad quotient" $\cX_\theta$ exists as a nice classical space
and is replaced by a noncommutative space, which
one can think of intuitively as a foliation by noncommutative tori. 
In fact, we have now the
domain of discontinuity replaced by the set   
$\Omega_\theta=\overline{H^3}\smallsetminus \{ z=0 \}$ on
which $\Gamma = e^{2\pi i \theta \Z}$ acts by $(z,y) \mapsto (e^{2\pi i \theta}z,y)$.
All the bad quotients $\{ (z,y): z\in \C^*,\, y=y_0 \}/\Gamma$ are (suspensions of) noncommutative
tori $\bT^2_\theta$ and the action of $\Gamma$ is trivial along the vertical direction,
so we can regard the resulting space $\cX_\theta$ as a one-parameter family of
noncommutative spaces $\bT^2_\theta\times \R^*_+$, one over 
each point $y$ in the vertical direction. More
precisely, one obtains in this description an algebra of coordinates that is
of the form $\cA_\theta \otimes C_0(\R^*_+)\otimes C_0(\R^*_+)$, where the
further commutative direction $\R^*_+$ corresponds to the vertical direction
in $\H^3$.  We can identify $C_0(\R^*_+)\otimes C_0(\R^*_+)$ with
$C_0(\R^*_+\times \R^*_+)$ and further $C_0(\R^2)$ by a homeomorphic
change of coordinates $\R^*_+\ni r=e^\rho$, $\rho\in \R$.  This non-unital
algebra $C_0(\R^2)$ has a unitization given by $C(S^2)$, by passing to the 
one point compactification.

\subsection{Noncommutative black holes}\label{ncblacksec}

We have described above a noncommutative deformation of the Euclidean
BTZ black hole at a purely topological level (as a $C^*$-algebra). We now
consider the metric aspect and the behavior under the Ricci flow.
First notice that the original BTZ black hole, as a finite quotient of $\H^3$ by
a discrete group of isometries, inherits from $\H^3$ the hyperbolic metric. On
an $n$-dimensional hyperbolic space $\H^n$, with the standard hyperbolic
metric,  the Ricci tensor satisfies $Ric(g)=-(n-1) g$, hence it flows as $g(t)=
(1+2(n-1)t) g$, expanding for all times, with a backward blowup at $t=-1/(2(n-1))$.

In the noncommutative deformation described above, one obtains different
behaviors under the Ricci flow, depending on how one treats the topology
of the commutative directions. As a product geometry $\bT^2_\theta\times \R^2$,
we can regard it as endowed with a product metric. In that case, the Ricci flow
preserves this form and each factor evolves with its own Ricci flow equation.
Thus, the noncommutative tori evolve according to the Ricci flow equation
we described in the previous section, while for the commutative direction we
can have, for example, shrinking spherical geometries, when compactifying to 
the one-point compactification $S^2$, steady torus geometries if compactifying
to an $S^1\times S^1$, or more interesting possibilities such as a Witten black
hole, also known as a Hamilton cigar soliton, see \cite{ChowLuNi} \S 4.3. 
This latter possibility is physically more
suitable for a black hole interpretation of the resulting space. 
A more interesting class of behaviors would 
be obtained if, instead of the product metric, one would consider a warped 
product, where the metric on the noncommutative torus $\bT^2_\theta$ (that
is, the element $k=e^{h/2}$ with $h=h^*\in \cA_\theta^\infty$) 
varies as a smooth function of the point in $\R^2$. 

\medskip

\section{Directions and perspectives}

The formulation of the Ricci flow we proposed here for 2-dimensional 
noncommutative tori is based on the behavior of eigenvalues of the
Laplacian on a 2-dimensional manifold, under the Ricci flow, assuming
some regularity conditions on the evolving eigenvalues and
eigenfunctions as in \cite{Cer07}. In the case of classical 2-dimensional
tori one knows \cite{Ham88} that an arbitrary initial metric flows to the
flat metric under the Ricci flow. The expression we obtained in
Theorem \ref{RicciRtau} appears to be, in its present form, too 
complicated to check directly whether this same property continues
to hold for the noncommutative tori, but we expect that further investigation
along these lines will answer this natural question.

In general, we hope that a good understanding of the Ricci flow on the 
noncommutative two torus would lead us to a
deeper understanding of the geometry of spectral triples and the underlying 
topology, as did the original Ricci flow for the classical case.

Whether a similar spectral formulation of the Ricci flow can be given for
a larger class of noncommutative spaces appears to be another interesting
question. The Ricci flow has been successfully
applied in understanding the relations between geometric and topological properties of Riemannian
manifolds, in particular 3-manifolds. A noncommutative analog of the Ricci flow would allow us to
pursue the same idea regarding this larger class of geometries.

In dimension higher than two, the equation obtained in \cite{Cer07}
for the Ricci flow (under the same strong assumptions) has
the more complicated form
$$ \frac{d\lambda}{dt} = \lambda \int_M f^2\, R\, d\mu - \int_M R\, |\nabla f|^2 d\mu +
2 \int_M {\rm Ric}(\nabla f, \nabla f)\, d\mu, $$
for the evolving Laplace eigenvalues and the corresponding normalized evolving eigenfunctions.
Thus, if one wishes to extend the approach proposed here, one needs to
provide a suitable interpretation of these additional terms, presumably in terms of  
the heat-kernel expansion of a suitable Laplacian operator on the noncommutative
geometry. A good model case on which to test this method in dimension three,
where the Ricci flow should be most interesting, may be the rich collection of
noncommutative 3-spheres, whose geometry and moduli spaces were studied
by Connes and Dubois-Violette in \cite{ConDub05}. 

Another possible point of view, which is naturally suggested by the case we
considered here, is the relation between the Ricci flow and the Yamabe flow
in 2-dimensions. This means that gaining a good understanding of how to
extend the Ricci flow to 2-dimensional noncommutative spaces, such as
the noncommutative tori, may also yield useful suggestions on how to extend
to noncommutative geometry the Yamabe flow and the Yamabe problem in
higher dimensions.  Also, by working with the determinant of the Laplacian
and its evolution equation under the Ricci flow, one would, by similar techniques,
be able to extend to noncommutative 2-tori the results of \cite{KotKor} and
obtain an analog of the Osgood--Phillips--Sarnak theorem in this setting,
\cite{OsPhiSa}.

In terms of other possible physical applications, it is well known (\cite{Frie},
see also \cite{Carfora} for a detailed discussion) 
that the Ricci flow arises as the weak coupling limit of
the renormalization group flow for non-linear sigma models. In recent years, 
generalizations of non-linear sigma models were considered, where the both
the target and the source space can be replaced by noncommutative spaces.
In particular, \cite{MatRos} gives a general form for an action functional where
the target space is a noncommutative Riemannian manifold (a spectral triple)
and the source space can also be a noncommutative space. In particular, they
study the case of sigma-models where these spaces are noncommutative
tori. It would be interesting to relate Ricci flow on noncommutative tori 
considered here to the renormalization group analysis of these sigma-models.

\bigskip

{\bf Acknowledgment.} This paper is based on the results of the first author's 
summer research project, within the Caltech program Summer Undergraduate 
Research Fellowships (SURF). The first author acknowledges support given from the bequest of Herbert J. Ryser (1923-1985)
through the Caltech mathematics department. The second author acknowledges support from NSF grants
DMS-0901221, DMS-1007207.

\bigskip

\section{Appendix}\label{appendix}

We report here, for the convenience of the reader, the explicit expression obtained
in \cite{ConTre09} for $S(\xi_1,\xi_2)$ in \eqref{intb2}, which we used in 
\S \ref{valuesec} above:
\begin{equation}\label{Sxi1xi2}
S(\xi_1,\xi_2) = 
\end{equation}
$  -b_0k\delta _1^2(k)-b_0k\delta _2^2(k)+\left(\xi_2 ^2+5 \xi_1
^2\right) \left(k^2 b_0^2\right)k\delta _1^2(k)+\left(5 \xi_2
^2+\xi_1 ^2\right) \left(k^2 b_0^2\right)k\delta _2^2(k)+2 \xi_2 ^2
\left(k^2 b_0^2\right)\delta _1(k)\delta _1(k)+6 \xi_1 ^2 \left(k^2
b_0^2\right)\delta _1(k)\delta _1(k)+\xi_2 ^2 \left(k^2
b_0^2\right)\delta _1^2(k)k+\xi_1 ^2 \left(k^2 b_0^2\right)\delta
_1^2(k)k+6 \xi_2 ^2 \left(k^2 b_0^2\right)\delta _2(k)\delta _2(k)+2
\xi_1 ^2 \left(k^2 b_0^2\right)\delta _2(k)\delta _2(k)+\xi_2 ^2
\left(k^2 b_0^2\right)\delta _2^2(k)k+\xi_1 ^2 \left(k^2
b_0^2\right)\delta _2^2(k)k-4 \xi_2 ^2 \xi_1 ^2 \left(k^4
b_0^3\right)k\delta _1^2(k)-4 \xi_1 ^4 \left(k^4 b_0^3\right)k\delta
_1^2(k)-4 \xi_2 ^4 \left(k^4 b_0^3\right)k\delta _2^2(k)-4 \xi_2 ^2
\xi_1 ^2 \left(k^4 b_0^3\right)k\delta _2^2(k)-8 \xi_2 ^2 \xi_1 ^2
\left(k^4 b_0^3\right)\delta _1(k)\delta _1(k)-8 \xi_1 ^4 \left(k^4
b_0^3\right)\delta _1(k)\delta _1(k)-4 \xi_2 ^2 \xi_1 ^2 \left(k^4
b_0^3\right)\delta _1^2(k)k-4 \xi_1 ^4 \left(k^4 b_0^3\right)\delta
_1^2(k)k-8 \xi_2 ^4 \left(k^4 b_0^3\right)\delta _2(k)\delta _2(k)-8
\xi_2 ^2 \xi_1 ^2 \left(k^4 b_0^3\right)\delta _2(k)\delta _2(k)-4
\xi_2 ^4 \left(k^4 b_0^3\right)\delta _2^2(k)k-4 \xi_2 ^2 \xi_1 ^2
\left(k^4 b_0^3\right)\delta _2^2(k)k+2 \xi_2 ^2 b_0k\delta
_1(k)b_0k\delta _1(k)+6 \xi_1 ^2 b_0k\delta _1(k)b_0k\delta _1(k)+2
\xi_2 ^2 b_0k\delta _1(k)b_0\delta _1(k)k+2 \xi_1 ^2 b_0k\delta
_1(k)b_0\delta _1(k)k-4 \xi_2 ^2 \xi_1 ^2 b_0k\delta _1(k)\left(k^2
b_0^2\right)k\delta _1(k)-4 \xi_1 ^4 b_0k\delta _1(k)\left(k^2
b_0^2\right)k\delta _1(k)-4 \xi_2 ^2 \xi_1 ^2 b_0k\delta
_1(k)\left(k^2 b_0^2\right)\delta _1(k)k-4 \xi_1 ^4 b_0k\delta
_1(k)\left(k^2 b_0^2\right)\delta _1(k)k+6 \xi_2 ^2 b_0k\delta
_2(k)b_0k\delta _2(k)+2 \xi_1 ^2 b_0k\delta _2(k)b_0k\delta _2(k)+2
\xi_2 ^2 b_0k\delta _2(k)b_0\delta _2(k)k+2 \xi_1 ^2 b_0k\delta
_2(k)b_0\delta _2(k)k-4 \xi_2 ^4 b_0k\delta _2(k)\left(k^2
b_0^2\right)k\delta _2(k)-4 \xi_2 ^2 \xi_1 ^2 b_0k\delta
_2(k)\left(k^2 b_0^2\right)k\delta _2(k)-4 \xi_2 ^4 b_0k\delta
_2(k)\left(k^2 b_0^2\right)\delta _2(k)k-4 \xi_2 ^2 \xi_1 ^2
b_0k\delta _2(k)\left(k^2 b_0^2\right)\delta _2(k)k-2 \xi_2 ^4
\left(k^2 b_0^2\right)k\delta _1(k)b_0k\delta _1(k)-16 \xi_2 ^2
\xi_1 ^2 \left(k^2 b_0^2\right)k\delta _1(k)b_0k\delta _1(k)-14
\xi_1 ^4 \left(k^2 b_0^2\right)k\delta _1(k)b_0k\delta _1(k)-2 \xi_2
^4 \left(k^2 b_0^2\right)k\delta _1(k)b_0\delta _1(k)k-12 \xi_2 ^2
\xi_1 ^2 \left(k^2 b_0^2\right)k\delta _1(k)b_0\delta _1(k)k-10
\xi_1 ^4 \left(k^2 b_0^2\right)k\delta _1(k)b_0\delta _1(k)k+4 \xi_2
^4 \xi_1 ^2 \left(k^2 b_0^2\right)k\delta _1(k)\left(k^2
b_0^2\right)k\delta _1(k)+8 \xi_2 ^2 \xi_1 ^4 \left(k^2
b_0^2\right)k\delta _1(k)\left(k^2 b_0^2\right)k\delta _1(k)+4 \xi_1
^6 \left(k^2 b_0^2\right)k\delta _1(k)\left(k^2 b_0^2\right)k\delta
_1(k)+4 \xi_2 ^4 \xi_1 ^2 \left(k^2 b_0^2\right)k\delta
_1(k)\left(k^2 b_0^2\right)\delta _1(k)k+8 \xi_2 ^2 \xi_1 ^4
\left(k^2 b_0^2\right)k\delta _1(k)\left(k^2 b_0^2\right)\delta
_1(k)k+4 \xi_1 ^6 \left(k^2 b_0^2\right)k\delta _1(k)\left(k^2
b_0^2\right)\delta _1(k)k-14 \xi_2 ^4 \left(k^2 b_0^2\right)k\delta
_2(k)b_0k\delta _2(k)-16 \xi_2 ^2 \xi_1 ^2 \left(k^2
b_0^2\right)k\delta _2(k)b_0k\delta _2(k)-2 \xi_1 ^4 \left(k^2
b_0^2\right)k\delta _2(k)b_0k\delta _2(k)-10 \xi_2 ^4 \left(k^2
b_0^2\right)k\delta _2(k)b_0\delta _2(k)k-12 \xi_2 ^2 \xi_1 ^2
\left(k^2 b_0^2\right)k\delta _2(k)b_0\delta _2(k)k-2 \xi_1 ^4
\left(k^2 b_0^2\right)k\delta _2(k)b_0\delta _2(k)k+4 \xi_2 ^6
\left(k^2 b_0^2\right)k\delta _2(k)\left(k^2 b_0^2\right)k\delta
_2(k)+8 \xi_2 ^4 \xi_1 ^2 \left(k^2 b_0^2\right)k\delta
_2(k)\left(k^2 b_0^2\right)k\delta _2(k)+4 \xi_2 ^2 \xi_1 ^4
\left(k^2 b_0^2\right)k\delta _2(k)\left(k^2 b_0^2\right)k\delta
_2(k)+4 \xi_2 ^6 \left(k^2 b_0^2\right)k\delta _2(k)\left(k^2
b_0^2\right)\delta _2(k)k+8 \xi_2 ^4 \xi_1 ^2 \left(k^2
b_0^2\right)k\delta _2(k)\left(k^2 b_0^2\right)\delta _2(k)k+4 \xi_2
^2 \xi_1 ^4 \left(k^2 b_0^2\right)k\delta _2(k)\left(k^2
b_0^2\right)\delta _2(k)k-2 \xi_2 ^4 \left(k^2 b_0^2\right)\delta
_1(k)kb_0k\delta _1(k)-12 \xi_2 ^2 \xi_1 ^2 \left(k^2
b_0^2\right)\delta _1(k)kb_0k\delta _1(k)-10 \xi_1 ^4 \left(k^2
b_0^2\right)\delta _1(k)kb_0k\delta _1(k)-2 \xi_2 ^4 \left(k^2
b_0^2\right)\delta _1(k)kb_0\delta _1(k)k-8 \xi_2 ^2 \xi_1 ^2
\left(k^2 b_0^2\right)\delta _1(k)kb_0\delta _1(k)k-6 \xi_1 ^4
\left(k^2 b_0^2\right)\delta _1(k)kb_0\delta _1(k)k+4 \xi_2 ^4 \xi_1
^2 \left(k^2 b_0^2\right)\delta _1(k)k\left(k^2 b_0^2\right)k\delta
_1(k)+8 \xi_2 ^2 \xi_1 ^4 \left(k^2 b_0^2\right)\delta
_1(k)k\left(k^2 b_0^2\right)k\delta _1(k)+4 \xi_1 ^6 \left(k^2
b_0^2\right)\delta _1(k)k\left(k^2 b_0^2\right)k\delta _1(k)+4 \xi_2
^4 \xi_1 ^2 \left(k^2 b_0^2\right)\delta _1(k)k\left(k^2
b_0^2\right)\delta _1(k)k+8 \xi_2 ^2 \xi_1 ^4 \left(k^2
b_0^2\right)\delta _1(k)k\left(k^2 b_0^2\right)\delta _1(k)k+4 \xi_1
^6 \left(k^2 b_0^2\right)\delta _1(k)k\left(k^2 b_0^2\right)\delta
_1(k)k-10 \xi_2 ^4 \left(k^2 b_0^2\right)\delta _2(k)kb_0k\delta
_2(k)-12 \xi_2 ^2 \xi_1 ^2 \left(k^2 b_0^2\right)\delta
_2(k)kb_0k\delta _2(k)-2 \xi_1 ^4 \left(k^2 b_0^2\right)\delta
_2(k)kb_0k\delta _2(k)-6 \xi_2 ^4 \left(k^2 b_0^2\right)\delta
_2(k)kb_0\delta _2(k)k-8 \xi_2 ^2 \xi_1 ^2 \left(k^2
b_0^2\right)\delta _2(k)kb_0\delta _2(k)k-2 \xi_1 ^4 \left(k^2
b_0^2\right)\delta _2(k)kb_0\delta _2(k)k+4 \xi_2 ^6 \left(k^2
b_0^2\right)\delta _2(k)k\left(k^2 b_0^2\right)k\delta _2(k)+8 \xi_2
^4 \xi_1 ^2 \left(k^2 b_0^2\right)\delta _2(k)k\left(k^2
b_0^2\right)k\delta _2(k)+4 \xi_2 ^2 \xi_1 ^4 \left(k^2
b_0^2\right)\delta _2(k)k\left(k^2 b_0^2\right)k\delta _2(k)+4 \xi_2
^6 \left(k^2 b_0^2\right)\delta _2(k)k\left(k^2 b_0^2\right)\delta
_2(k)k+8 \xi_2 ^4 \xi_1 ^2 \left(k^2 b_0^2\right)\delta
_2(k)k\left(k^2 b_0^2\right)\delta _2(k)k+4 \xi_2 ^2 \xi_1 ^4
\left(k^2 b_0^2\right)\delta _2(k)k\left(k^2 b_0^2\right)\delta
_2(k)k+8 \xi_2 ^4 \xi_1 ^2 \left(k^4 b_0^3\right)k\delta
_1(k)b_0k\delta _1(k)+16 \xi_2 ^2 \xi_1 ^4 \left(k^4
b_0^3\right)k\delta _1(k)b_0k\delta _1(k)+8 \xi_1 ^6 \left(k^4
b_0^3\right)k\delta _1(k)b_0k\delta _1(k)+8 \xi_2 ^4 \xi_1 ^2
\left(k^4 b_0^3\right)k\delta _1(k)b_0\delta _1(k)k+16 \xi_2 ^2
\xi_1 ^4 \left(k^4 b_0^3\right)k\delta _1(k)b_0\delta _1(k)k+8 \xi_1
^6 \left(k^4 b_0^3\right)k\delta _1(k)b_0\delta _1(k)k+8 \xi_2 ^6
\left(k^4 b_0^3\right)k\delta _2(k)b_0k\delta _2(k)+16 \xi_2 ^4
\xi_1 ^2 \left(k^4 b_0^3\right)k\delta _2(k)b_0k\delta _2(k)+8 \xi_2
^2 \xi_1 ^4 \left(k^4 b_0^3\right)k\delta _2(k)b_0k\delta _2(k)+8
\xi_2 ^6 \left(k^4 b_0^3\right)k\delta _2(k)b_0\delta _2(k)k+16
\xi_2 ^4 \xi_1 ^2 \left(k^4 b_0^3\right)k\delta _2(k)b_0\delta
_2(k)k+8 \xi_2 ^2 \xi_1 ^4 \left(k^4 b_0^3\right)k\delta
_2(k)b_0\delta _2(k)k+8 \xi_2 ^4 \xi_1 ^2 \left(k^4
b_0^3\right)\delta _1(k)kb_0k\delta _1(k)+16 \xi_2 ^2 \xi_1 ^4
\left(k^4 b_0^3\right)\delta _1(k)kb_0k\delta _1(k)+8 \xi_1 ^6
\left(k^4 b_0^3\right)\delta _1(k)kb_0k\delta _1(k)+8 \xi_2 ^4 \xi_1
^2 \left(k^4 b_0^3\right)\delta _1(k)kb_0\delta _1(k)k+16 \xi_2 ^2
\xi_1 ^4 \left(k^4 b_0^3\right)\delta _1(k)kb_0\delta _1(k)k+8 \xi_1
^6 \left(k^4 b_0^3\right)\delta _1(k)kb_0\delta _1(k)k+8 \xi_2 ^6
\left(k^4 b_0^3\right)\delta _2(k)kb_0k\delta _2(k)+16 \xi_2 ^4
\xi_1 ^2 \left(k^4 b_0^3\right)\delta _2(k)kb_0k\delta _2(k)+8 \xi_2
^2 \xi_1 ^4 \left(k^4 b_0^3\right)\delta _2(k)kb_0k\delta _2(k)+8
\xi_2 ^6 \left(k^4 b_0^3\right)\delta _2(k)kb_0\delta _2(k)k+16
\xi_2 ^4 \xi_1 ^2 \left(k^4 b_0^3\right)\delta _2(k)kb_0\delta
_2(k)k+8 \xi_2 ^2 \xi_1 ^4 \left(k^4 b_0^3\right)\delta
_2(k)kb_0\delta _2(k)k$.

\end{document}